\documentclass[article,twocolumn,superscriptaddress]{revtex4-1}
\usepackage{graphicx, tabularx, amsmath}
\usepackage{hyperref}
\begin{document}
\title[Student understanding of Coriolis...]{Improving students' understanding of rotating frames of reference using videos from different perspectives}

\author{S. K\"uchemann}\email{s.kuechemann@physik.uni-kl.de}
\affiliation{Department of Physics, Physics Education Research Group, Technische Universit\"at Kaiserslautern, Erwin-Schr\"odinger-Str. 46, 67663 Kaiserslautern, Germany}

\author{P. Klein}
\affiliation{Department of Physics, Physics Education Research Group, Technische Universit\"at Kaiserslautern, Erwin-Schr\"odinger-Str. 46, 67663 Kaiserslautern, Germany}

\author{H. Fouckhardt}
\affiliation{Department of Physics, Integrated Optoelectronics and Microoptics Research Group, Technische Universit\"at Kaiserslautern, Erwin-Schr\"odinger-Str. 46, 67663 Kaiserslautern, Germany}

\author{S. Gr\"ober}
\affiliation{Department of Physics, Physics Education Research Group, Technische Universit\"at Kaiserslautern, Erwin-Schr\"odinger-Str. 46, 67663 Kaiserslautern, Germany}

\author{J. Kuhn}
\affiliation{Department of Physics, Physics Education Research Group, Technische Universit\"at Kaiserslautern, Erwin-Schr\"odinger-Str. 46, 67663 Kaiserslautern, Germany}

\begin{abstract}
The concepts of the Coriolis and the centrifugal force are essential in various scientific fields and they are standard components of introductory physics lectures. In this paper we explore how students understand and apply concepts of rotating frames of reference in the context of an exemplary lecture demonstration experiment. We found in a $Predict-Observe-Explain$-setting, that after predicting the outcome prior to the demonstration, only one out of five physics students correctly reported the observation of the trajectory of a sphere rolling over a rotating disc. Despite this low score, a detailed analysis of distractors revealed significant conceptual learning during the observation of the experiment. In this context, we identified three main misconceptions and learning difficulties. First, the centrifugal force is only required to describe the trajectory if the object is coupled to the rotating system. Second, inertial forces cause a reaction of an object on which they act. And third, students systematically mix-up the trajectories in the stationary and the rotating frame of reference. Furthermore, we captured students' eye movements during the $Predict$ task and found that physics students with low confidence ratings focused longer on relevant task areas than confident students despite having a comparable score. Consequently, this metric is a helpful tool for the identification of misconceptions using eye tracking. Overall, the results help to understand the complexity of concept learning from demonstration experiments and provide important implications for instructional design of introductions to rotating frames of reference.  
\end{abstract}

\maketitle

\section{Introduction}
Rotating frames of reference play an important role in a variety of fields in physics. Accordingly, Coriolis and centrifugal terms need to be considered for an accurate account of the theoretical description. While the Coriolis force was originally introduced to describe the energy transfer in waterwheels, nowadays it is applied to problems in meteorology \cite{Holton,Persson1}, oceanography \cite{Kirby}, astrophysics \cite{Dintrans}, optics \cite{Bliokh} and nuclear physics \cite{Walker}. Given this wide range of applications, the Coriolis and the inertial centrifugal force are common topics in introductory physics courses in college-level education and, accordingly, there is a large number of experiments and online materials \cite{youtubeNatGeo} which intend to demonstrate the Coriolis effect, i.e. the apparent deflection of an object by the Coriolis force. However, there are several shortcomings and false accounts, outlined below, potentially causing misconceptions and complications in students' understanding.\\ 
In this paper we explore how students understand and apply concepts of rotating frames of reference in the direct context of an exemplary lecture demonstration experiment. Therefore, we identify and study relevant misconceptions, the non-obvious learning effect of experiment observation and the relationship between response security and duration of focus on relevant areas (as measured by eye tracking).\\
The paper is structured in the following way. After this introduction, an overview of the state of the art and the preliminary work follows in the second chapter and the third part explains the materials and methods used in this work. The subsequent section contains the results of the Predict-Observe-Explain test, including self-confidence ratings, student interviews and eye-tracking data in the context of an exemplary demonstration experiment of rotating frames of reference. Then, these results are discussed in the context of previous literature and, eventually in the final chapter, we conclude the manuscript with the main consequences of the results for physics education research.
\section{State of the art and preliminary work}
\subsection{Simplified conceptions of the Coriolis effect and the centrifugal force}
In simplified depictions of a curved trajectory of an object in a rotating frame of reference (RFR), the Coriolis force is often presented as the cause for the deflection. However, according to Eq. (\ref{eq:accel}) (see Appendix), the inertial centrifugal force (ICF) also acts on the object in a vector sum with the Coriolis force. The fact that the inertial centrifugal force is a necessary quantity to describe the trajectory of an object in a rotating frame of reference can be understood from two arguments of a thought experiment.\\ 
(A) Let us imagine a two-dimensional case where a plane flies in a uniform motion over a large rotating disc starting from the center of rotation. If an observer located on the disc only used the Coriolis force for the description of the curved trajectory, he/she would expect that the plane returns to the center of rotation at some point in time because the Coriolis force is always perpendicular to the direction of motion thus leading to a circular trajectory. For an observer in a stationary frame of reference (SFR), however, it is obvious that this case would not occur because the plane flies in a uniform motion due to the absence of any real force. In reality, the plane would pursue a spiral trajectory for the observer in the RFR which is the consequence of the vector sum of the inertial centrifugal force and Coriolis force.\\ 
(B) During the aforementioned motion of the plane, the absolute value of the velocity $\left|\vec v'\right|$ in the RFR would increase according to Eq. (\ref{eq:vel}) because the absolute value of the velocity in the SFR $\left|\vec v\right|$ and the angular velocity are constant and $\left|\vec r\right|$ increases. Since the Coriolis force is always perpendicular to the direction of motion, it cannot be the reason for this apparent increase in $\left|\vec v'\right|$. Only the centrifugal force which points outwards from the center of rotation can be responsible for this effect.\\
The misconception that the centrifugal force only occurs when the object is somehow coupled (e.g. by friction or a rope) to the rotating system \cite{Persson2} is potentially guided by empirical experiences, such as the feeling of a force pointing outwards. Here, the underlying problem is that a person when sitting in a carousel or in a car driving through a turn actually feels a force acting on him or her, i.e. the body reacts to the force, because the person is actually partially coupled to the rotating system. This seems to be in conflict with the aforementioned characteristics of a virtual force. This cognitive dissonance can be resolved by discriminating between the centrifugal force which occurs as a reaction to a Centripetal force (here termed "Reactive centrifugal force", RCF) and the one which occurs as a virtual force in a RFR ("inertial centrifugal force") \cite{Kobayashi}. Sometimes text books and scientific articles lack this helpful linguistic distinction \cite{Corben, Demtroeder, Kibble, Stommel, Persson2}. The reason for this could be that the mathematical equations are the same, only the situations in which they occur in and how they are perceived are different. For the occurrence of the RCF, a coupling to the RFR is indeed required, for the occurrence of the ICF it is not. Accordingly, the RCF can be perceived when driving through a turn or sitting in a merry-go-round while the occurrence of the ICF, for instance for a passenger in an air plane which flies in a uniform motion over a rotating disc, cannot be felt.  
\subsection{Experimental lecture demonstrations and students' understanding of the Coriolis force}
Lecture demonstrations in the classical sense mean the demonstrations of experiments by the lecturer during the class while the students passively observe the presentation. The intention of the lecturer is often that the students process the information and understand their observations by integrating it into their concept knowledge \cite{Miller}. Unfortunately, despite their regular use in introductory physics lectures, it has been shown that demonstrations will have little effect on students' concept learning if the students passively observe the experiments \cite{Crouch}. At the same time, the correct observation of a lecture demonstration is a necessary prerequisite for concept learning \cite{Miller}.\\
In this context, Predict-Observe-Explain (POE) is an interactive teaching scenario which can be implemented during experimental lecture demonstrations \cite{White,Champagne,Sokoloff}. While it is sometimes proposed as an eight-step approach, here we reduce it to three central steps \cite{Gunstone2,Treagust}. First, in the "Predict" phase, the students are asked to make an educated guess of the outcome of the experiment. This step helps to initiate learning processes by reflecting on and relating to theoretical backgrounds and thus forming a mental model which links the theory to the experiment \cite{Treagust}. In the second, "Observe" phase the experiment is demonstrated and the students visually perceive its process and outcome. Here, students are expected to relate their observation to the previously anticipated result and, consequently, approve or reconsider their mental model \cite{Champagne}. In the final "Explain" phase the outcome of the experiment is revisited, typically by the teacher. In this part the teacher explains the established link between the theory and the outcome of the experiment.\\
Previous research on the impact of POE scenarios have shown that it is more effective for students' learning and students pay more attention than during classical lecture demonstrations in which the instructor only performs the experiment and explains it in the framework of the established theory. \\
To our knowledge there is no quantitative study which examines students' misconceptions of Coriolis force and inertial centrifugal force. Stommel et al. report that students consider the Coriolis effect as "mysterious" and a result of "formal mathematical manipulations" as pointed out by Persson \cite{Stommel,Persson2}. Previously observed misconceptions of students in mechanics imply, for instance, the "Motion implies a force" misconception \cite{clement, Gunstone}. This misconception potentially still persist in the students' understanding, thus complicating the students' conceptional learning of rotating frames of references and may translate to our study. We have accounted for these potential misconceptions in the posttest.
\subsection{Analysis methods}
\subsubsection{Eye-tracking in educational research}
During the POE tasks and the instruction (between $Observe$ and $Explain$) we have recorded the motion of the eyes of the students. In general, the eye-tracking technique has gained growing attention in educational research in the past years, since several cognition-psychological and educational questions can be addressed with this method. Theoretically, the eye-tracking technique is founded on the eye-mind hypothesis, which means that the visual focus is located on information which are cognitively processed. This hypothesis was originally formulated by Just and Carpenter \cite{Just} and was later on experimentally confirmed by Kustov and Robinson \cite{Kustov}. Therefore, eye-tracking can be used, for instance, to validate prevalent cognitive and multimedia theories \cite{VanGog}, to reveal student strategies during problem-solving \cite{Susac}, and to discriminate between expert and novice eye-gaze patterns \cite{Ericsson1}, thus leading to improved instructional designs \cite{Pascal1}. \\
In this context, Gegenfurtner et al concluded in a meta-analysis that experts, in comparison to non-experts, have shorter fixation duration but more fixations on relevant areas and longer saccades \cite{Gegenfurtner}, confirming a number of theories, such as the theory of long-term working memory \cite{Ericsson2} and the information-reduction hypothesis \cite{Haider}.\\
In the context of physics education, eye-tracking has been used, for instance, in the context of kinematic graphs and vector fields. Klein et al. found that high-performing students rather follow an equation-based approach than low-performers, thus executing more vertical and horizontal eye-movements during the interpretation of two-dimensional vector fields \cite{Pascal1}. Apart from that, Madsen et al found that the response accuracy is correlated to focus duration on relevant areas \cite{Madsen1}. To our knowledge, eye-tracking has not been applied in the context of demonstration experiments in POE settings. 
\subsubsection{Self-confidence ratings}
In this study, we use self-confidence ratings after the students have answered a question. These meta-cognitive ratings in a single choice format reflect the ability of students to self-monitor their thought processes, which comprises a reflection of the understanding of the topic and the performance in the task \cite{Sharma}. In common interpretations of confidence ratings, the difference between the confidence rating and the accuracy of the response is termed \textit{bias}. The bias is low for a student who has a comparable confidence rating to his or her accuracy and, consequently, it would be high if the student tends to over- or underestimate his or her performance. The level of the bias is an indication for the \textit{calibration}. A deviation from a zero bias is an indication for a lack of calibration. The relatively robust effect of overconfidence can be explained within the probabilistic mental model (PMM) theory proposed by Gigerenzer et al. \cite{Gigerenzer,Fischhoff}, in which confidence judgments are first a spontaneous consequence of a local mental model (LMM). In case, where a LMM in the context of a specific task fails, a PMM is created in which the person retrieves probabilistic cues from the environment. The mismatch between the cue validity and ecological validity, which is the true account of a certain situation, might be one of two reasons for an overconfidence. The second potential reason within the PMM theory is that the set of information retrieved from the environment is not a representative selection for the reference set \cite{Gigerenzer} and, for comparison, the reason is not a misled perception of the task difficulty \cite{Allwood,May}.\\
In the field of physics education, Planinic et al. found significantly higher confidence ratings for wrong answers in the area of Newtonian dynamics than in the area of  electrical circuits, suggesting that concepts of Newtonian dynamics are more prone to misconceptions \cite{Planinic}.\\
In this work, we use the confidence ratings as an aid to identify underlying misconceptions, which reveal themselves when the student appears to be rather confident with an incorrect answer. Furthermore, this study explores the influence of the level of calibration on the conceptual learning within a POE setting and relates the confidence to eye-tracking metrics. \\
Within this educational framework we formulate three research questions:
\begin{itemize}
\item What are the prevailing misconceptions of physics students in the field of rotating frames of reference? 
\item What is the influence of the demonstration experiment on learning about the outcome of the experiment?
\item	Is there a specific eye-movement pattern which relates to the performance or confidence of physics students within a POE setting?
\end{itemize}
\section{Materials and methods}
\subsection{Participants}
The sample consists of 21 freshman students with a physics major of the Technische Universit\"at Kaiserslautern, Germany. The students were participants of the lecture "Experimental Physics 1" where they had seen experimental demonstrations and the mathematical derivation of the topic of rotating frames of reference in one lecture, one tutorial, one problem sheet, and one recitation session prior to participation in this study. Participation in this study was voluntary and was compensated with 10 Euro. The study took place several weeks before the final exam of the lecture and the students expected that the topic might be part of the exam.
\subsection{Experimental setup}
The setup consists of a rotating disc with a diameter of 55 cm which is connected to a motor that allows the disc to rotate at a constant angular velocity (see Fig. \ref{Fig1_Design_Setup}a and \ref{Fig1_Design_Setup}b). Initially, the sphere rests at the end of a tilted rail which is attached to the rotating disc pointing in the direction of the center of rotation (since the rail is attached to the rotating disc the sphere receives an initial tangential velocity component). As soon as the rail passes a trigger, the sphere starts to roll down the rail (from this acceleration the sphere receives an initial radial velocity component). The experiment is recorded from the top via two cameras. The first camera is connected to the stationary frame and does not move while the disc rotates. The second camera is attached to the rotating disc, allowing the observation in the perspective of a rotating observer. \\
In the stationary system, the sphere moves uniformly in a straight line on the left side of the disc in respect to the center of rotation. Note here, that it does not run through the center because the resulting motion is a superposition of the tangential and the radial part. This means that the answer (b) is correct in the stationary frame of reference $K$ (see Fig. \ref{Fig1_Design_Setup}d). In the rotating frame of reference, the trajectory d) describes the motion correctly (see Fig. \ref{Fig1_Design_Setup}d).   
\subsection{Study design}
\begin{figure}[ht!]
	\includegraphics[width=\linewidth]{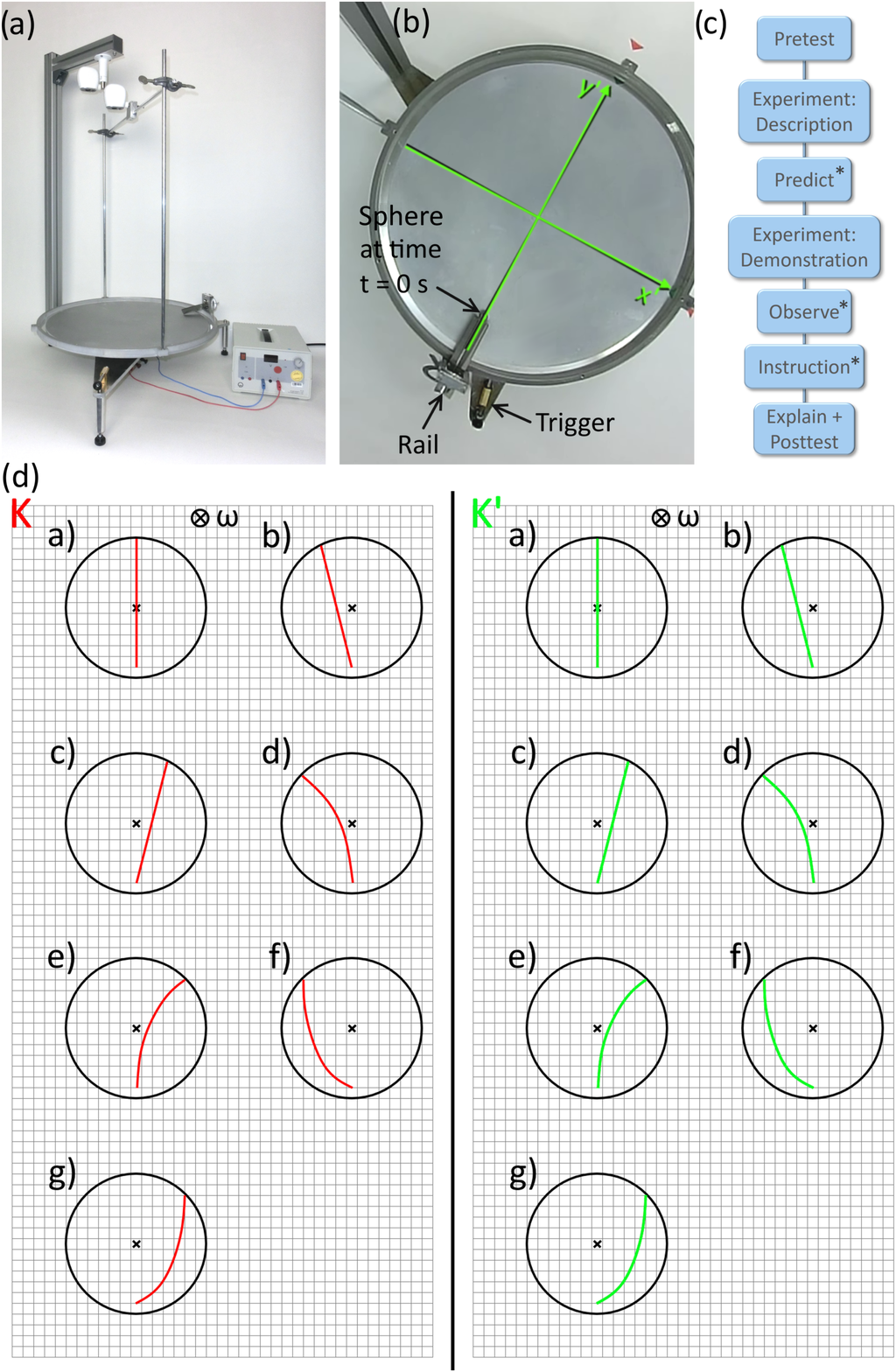}
	\caption{(a) The experimental setup for the demonstration of rotating frames of references. (b) Top view of the rotating disc . (c) Study design where $^*$ indicates the parts in which the eye movements were recorded and (d) the alternative answers of the POE items in the stationary frame of reference $K$ and the rotating frame of reference $K^{\prime}$. In both coordinate systems the distractors are identical.}
	\label{Fig1_Design_Setup}
\end{figure}
The study design is outlined in Fig. \ref{Fig1_Design_Setup}c. The pretest consisted of three single choice items in a paper-pencil test assessing the understanding of essential representations of vectors. Thus, we verified whether or not the students had visual understanding of typical depictions of rotating frames of reference used in this study - a necessary prerequisite for learning from multiple visual representations as in the instruction part \cite{Rau1}. It was followed by an explanation of the experimental setup and the procedure of the experiment (without demonstration yet) by the instructor (see Fig. \ref{Fig1_Design_Setup} b,c). In this phase, the students were allowed to ask questions.\\ 
Afterwards, the students were asked two questions to anticipate the trajectory of the sphere in a stationary frame of reference (first) as well as in a rotating frame of reference (second), each of them in a single choice format (see Fig. \ref{Fig1_Design_Setup}d for answer alternatives), which is termed the $Predict$ phase. These two questions were computer-based and the eye movements were recorded. After each prediction, the students were asked to rate their confidence on a four-point Liekert scale ranging from ``very confident" to ``very unconfident".\\
Then, the instructor demonstrated the experiment twice (part $Observe$). The students were allowed to observe it from every perspective. This part was supposed to closely resemble an ideal situation of a lecture demonstration. Then, the students were asked to answer the same two questions as in the $Predict$ phase in order to report their observation of the trajectory in the stationary and in the rotating frame of reference. Again, we used eye-tracking and confidence ratings for these two computer-based items.\\
Subsequently, the students received the computer-based instruction consisting of two text pages and six videos. The first page displayed a standard text book instruction of inertial forces including the equations of the Coriolis and centrifugal force. The second page explained the trajectory of the sphere rolling over a rotating disc in the particular context of the previously demonstrated experiment. This page also contained two snapshots of the final location of the sphere during the experiment from each perspective (see Fig. 2) augmented with circles and arrows indicating the trajectory and velocity vectors in both frames of reference.\\
\begin{figure}[ht!]
	\includegraphics[width=\linewidth]{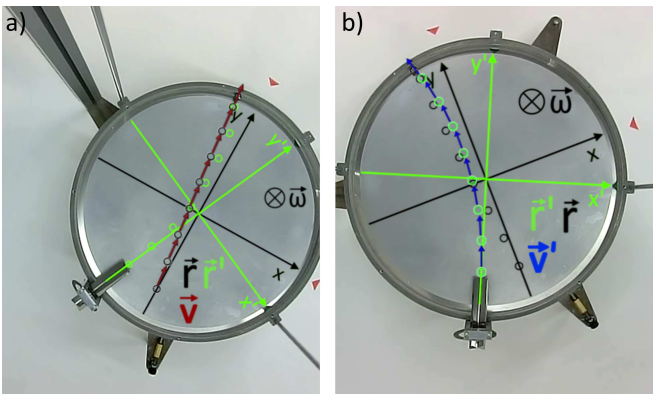}
	\caption{Snapshots during the final phase of the experiment in the stationary frame of reference (a) and the rotating frame of reference (b). The black coordinate system $K$ (axes x and y) is stationary and the green coordinate system $K^{\prime}$ (axes x$^{\prime}$ and y$^{\prime}$) rotates at the same angular velocity as the disc.}
	\label{Fig2_Snap}
\end{figure}
After this first instruction page, three videos from each of the two perspectives (i.e. six videos in total) were shown to the students. The first video showed the experiment recorded by the stationary camera in real time. It was augmented with the same information as in the snapshots in Fig. \ref{Fig2_Snap}. The two following videos were identical to the first one but they were played in slow motion ($4 \times$ slower). The three videos recorded from the rotating camera were produced in the same format and played in the same order. The students had no option to pause or replay the videos.\\
After the instruction, the posttest in a paper-pencil format and two computer-based questions followed. It consisted of seven true-false items, two items $Explain$ (identical to $Predict$ and $Observe$) and seven single choice items, two of which had a direct link to the experiment and they were posed in the Eye-tracking setting. After completing the posttest, the students were asked to comment on their responses of two single choice items from the posttest in an audio interview. The aim of the interview was to reveal potential misconceptions. Therefore, these two questions were directly motivated by the misleading depictions in literature (see above).
\subsection{Eye-tracking equipment}
The motions of the eyes were recorded using a Tobii X3-120 stationary eye-tracking system with a visual-angle resolution of $0.40^{\circ}$ and a sampling rate of 120~Hz. The questions were presented on a 22-inch computer screen with a resolution of $1920\times 1080$ pixels and refresh rate of 75 Hz. The eye-tracking system was operated and the data was partially analyzed using the software Tobii Studio.
\section{Results}
\subsection{Test scores of POE items}
\begin{figure}[ht!]
	\includegraphics[width=\linewidth]{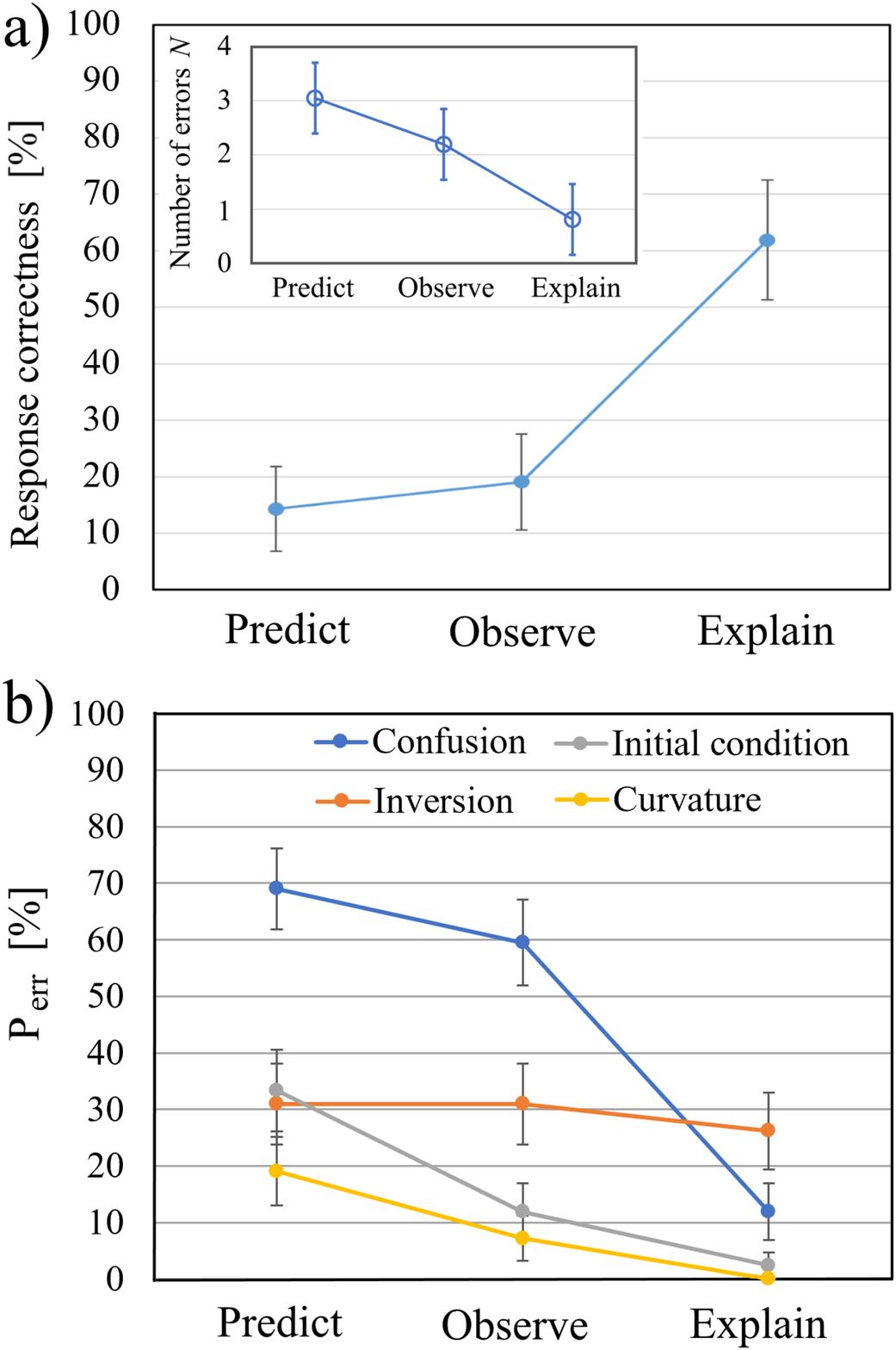}
	\caption{Test scores of POE items (a) and probability of different error types (b). The inset in panel (a) shows the total number of errors $N$ of the POE items which refers to the analysis in Sec. \ref{Sec:AoDistr} (see also Tab. \ref{tab:distr})}
	\label{Fig3_scores}
\end{figure}

The test scores of the POE questions are shown in Fig. \ref{Fig3_scores}a. The score in each POE part is the average score from two questions about the trajectory of the sphere on the rotating disc: The first question is about the trajectory in a stationary coordinate system and the second one is about the trajectory in a rotating coordinate system. \\
It is noticeable that the students have very low scores during the $Predict$ phase. The demonstration of the experiment which is the only intervention between $Predict$ and $Observe$ has no significant effect on the score ($p=0.27$), which means that 80 \% of the students are unable to report the trajectory of a sphere in two coordinate systems after observing the experiment twice. After the instruction including a theoretical textbook introduction, two augmented photographs (see Fig. \ref{Fig2_Snap}) and six videos, in the part $Explain$, 62 \% of the students report the trajectories of the sphere correctly.\\
The average confidence ratings do not change between $Predict$ ($2.2 \pm 0.7$, where 1 $=$ very confident and 4 $=$ very unconfident) and $Observe$ ($2.1 \pm 0.8$).
\subsection{Analysis of distractors}\label{Sec:AoDistr}
For a deeper understanding of error sources and the influence of interventions, we divided the distractors of the POE items into different types. The distractors are displayed in Fig. \ref{Fig1_Design_Setup}d. The following example may demonstrate the motivation for this process. A student who chooses a straight trajectory through the center of rotation in the stationary frame of reference $K$ has a different perception of the trajectory and a potentially different concept of the situation than a student who chooses a curved trajectory which is deflected to the right of the center in $K$, despite the fact that both answers are incorrect.\\ 
In detail, we identified four different types of errors among the distractors of the POE task: \\
\begin{enumerate}
\item[I: ] Confusion of $K$ and $K^{\prime}$: When a student chooses a curved trajectory in $K$ or a straight trajectory in $K^{\prime}$. In $K$, this error type occurs when a student chooses either one of the distractors (d), (e), (f) or (g). In $K^{\prime}$ this error type occurs when a student chooses either one of distractors (a), (b), (c). 
\item[II: ] Inversion: Here, the student chooses a distractor which depicts a trajectory to the right in respect to the center of rotation. Included distractors in $K$: (c), (e), (g). Included distractors in $K^{\prime}$: (c), (e), (g).
\item[III: ] Initial condition: Here, the student does not consider that the sphere also has a tangential velocity component in $K$ and chooses the trajectory through center of rotation. Included distractors in $K$: (a). Included distractors in $K^{\prime}$: (a).
\item[IV: ] Curvature: In this error, the student selects a distractor with an incorrect curvature. Included distractors in $K$: (f), (g). Included distractors in $K^{\prime}$: (f), (g).
\end{enumerate}
Following this line of thought, the assignment the error types to the different distractors of the POE items implies that some distractors exhibit more than one error (see Tab. \ref{tab:distr}). The number $N$ of errors for one answer alternative ranges from $0-3$ in $K$ and $0-2$ in $K^{\prime}$.
\begin{table}[ht]
	\centering
		\caption{Error type (ET) and number of errors ($N$) for each answer alternative in $K$ and $K^{\prime}$.}
		\label{tab:distr}
		\begin{tabular}{lcccccccc}
		\hline
		Distractor  &$~~~$& $ET_K$ &$~~~$& $N_{K}$ &$~~~$& $ET_{K^{\prime}}$ &$~~~$& $N_{K^{\prime}}$ \\ \hline 
		$a$ &$~~~$& III &$~~~$& 1 &$~~~$& I, III &$~~~$& 2 \\ 
		$b$ &$~~~$& cor &$~~~$& 0 &$~~~$& I &$~~~$& 1 \\ 
		$c$ &$~~~$& II &$~~~$& 1 &$~~~$& I,II &$~~~$& 2 \\ 
		$d$ &$~~~$& I &$~~~$& 1 &$~~~$& cor &$~~~$& 0 \\ 
		$e$ &$~~~$& I,II &$~~~$& 2 &$~~~$& II &$~~~$& 1 \\ 
		$f$ &$~~~$& I,IV &$~~~$& 2 &$~~~$& IV &$~~~$& 1 \\ 
		$g$ &$~~~$& I,II,IV &$~~~$& 3 &$~~~$& II,IV &$~~~$& 2 \\ \hline
		\end{tabular}
\end{table}
Fig. \ref{Fig3_scores}b shows the probability of each error during the POE tasks. The error probability displayed in this figure is the average probability of the ones within the two coordinate systems. It is noticeable that, as a consequence of the experiment demonstration, all errors are reduced between $Predict$ and $Observe$ except the inversion error. In fact, the average difference between the total number of errors in $Predict$ ($N=3.05 \pm 1.40$) and $Observe$ ($N=2.19 \pm 1.29$) exhibits a significant medium effect (Cohen's $d=0.64$, $p<0.05$, see inset of Fig. \ref{Fig3_scores}a). In comparison, the average difference between the number of errors in $Observe$ and $Explain$ ($N=0.81 \pm 1.03$) exhibits a significant very large effect (Cohen's $d=1.18$, $p<0.001$). The largest improvement between $Predict$ and $Observe$ was found in the observation of the ``initial condition", i.e. the students were able to correct their prediction that the sphere does not go through the center of rotation. In contrast, the observation of the demonstration experiment did not affect the ``Inversion"-error, i.e. the students did not recognize that the sphere had been deflected to the left side in respect to the center of rotation if they had previously predicted that the trajectory is located on the right side of the disc.\\
The largest improvement between $Observe$ and $Explain$, as a consequence of the instruction, is the ``Confusion"-error, i.e. after the instruction most students can relate a linear trajectory to the stationary frame of reference and a curved trajectory to the rotating frame of reference. And again, as previously observed between $Predict$ and $Observe$, the instruction did not affect the ``Inversion"-error, i.e. even after seeing the trajectory in an augmented photograph and in six videos those students who previously made the ``Inversion"-error still fail to realize that the trajectory in both coordinate systems is located on the left side of the disc. Apart from this, the ``Curvature"-error could be corrected entirely using the instruction and also the error in the ``initial condition" was made only by one student after the instruction.\\
\begin{table}[ht]
	\centering
		\caption{Average confidence ratings in respect to error type in the $Predict$ and the $Observe$ part. The numerical confidence values correspond to: 1 $=$ very confident, 2 $=$ confident, 3 $=$ unconfident and 4 $=$ very unconfident.}
		\label{tab:confis}
		\begin{tabular}{lcccc}
		\hline
		Error  &$~~~$& $Predict$ &$~~~$& $Observe$ \\ 
		 type &$~~~$& Confidence &$~~~$& Confidence \\ \hline 
		I &$~~~$& 2.2 &$~~~$& 2.2 \\ 
		II &$~~~$& 2.6 &$~~~$& 2.4 \\ 
		III &$~~~$& 1.9 &$~~~$& 1.2 \\ 
		IV &$~~~$& 2.5 &$~~~$& 3.3  \\ \hline
		\end{tabular}
\end{table}
Tab. \ref{tab:confis} shows the confidence ratings of each error type during the $Predict$ and $Observe$ questions. It is noticeable that the confidence ratings between $Predict$ and $Observe$ are very similar and no significant change can be observed. The students which made an error of type III had the highest confidence, particularly in the $Observe$ part. In contrast, the students which made an error of type IV had the lowest confidence in the $Observe$ part. \\
The confidence ratings neither correlate with the number of errors ($r_{\textrm{Pearson}}=0.29$, $p=0.90$) nor with the improvement between the $Predict$ and the $Observe$ part ($r_\textrm{Pearson}=-0.35$, $p=0.06$). Apart from this, there is no significant difference between the confidence levels of answers to questions regarding the trajectory in the stationary frame of reference $K$ and the rotating frame of reference $K^{\prime}$. \\
In the context of confidence ratings in science education research, the Dunning-Kruger effect, which states that low-performing students rather tend to overestimate their performance, is often reported \cite{Kruger,Lindsey}. Here, we are not able to verify this effect due to the small number of correct answers during the $Predict$ and the $Observe$ part.
\subsection{Student interviews}
The confidence ratings suggested that there are small differences between error types. To consolidate this finding and to identify misconceptions we performed student interviews after completing the posttest. In these interviews we asked the students to comment on their answers of two particular questions Q1 and Q2 of the posttest. In question Q1 the students were asked to name the forces, which are required to describe the trajectory of an airplane flying from the center of an rotating disc outwards in a uniform motion. In question Q2 the students were asked to predict whether or not water would slosh over the edge of a glass if the glass moves along a curved trajectory in $K^{\prime}$ but uniformly along a straight line in $K$. The two questions and the possible answers are outlined in the Appendix. The interviews were conducted in German and, afterward, translated to English. Language errors were corrected to improve readability. \\
\begin{table}[ht]
	\centering
		\caption{Distribution of answers of the two interview questions Q1 and Q2 (see Appendix for the questions and possible answers). The correct answer is marked with a dagger.}
		\label{tab:Interv}
		\begin{tabular}{lcccc}
		\hline
		  Distractor &$~~~$& Q1 [\%] &$~~~$& Q2 [\%]\\ \hline
		a) &$~~~$& 28.6$^{\dagger}$ &$~~~$& 0 \\ 
		b) &$~~~$ & 0 &$~~~$& 4.8 \\ 
		c) & $~~~$& 42.8 & $~~~$& 14.3 \\ 
		d) & $~~~$& 28.6 &$~~~$ & 4.8  \\ 
		e) & $~~~$& - & $~~~$& 23.8\\
		f) & $~~~$& - &$~~~$ & 52.8$^{\dagger}$\\ \hline
		\end{tabular}
\end{table}
Table \ref{tab:Interv} shows the probability of each possible answer. In question Q1, the distractor (c) has the highest probability. This corresponds to the answer that only the Coriolis force is required to describe the trajectory of the airplane flying over a rotating disc. In the interviews, all students who chose this answer either justified their response by the thought that the airplane has no connection to the rotating disc or argued that in the absence of a centripetal force, no centrifugal one is required for the description of the trajectory. Here, we show two examples of medium performing students M1 and M2 and one example of a high performing student H:\\
Instructor:\textit{``Please comment on your answer of question Q1."}\\
M1: \textit{``I ticked the third one, because actually only the Coriolis force would have to work. I originally assumed that the Coriolis force is a counterforce of the centrifugal force. But since this is wrong and actually the centrifugal force is the counterforce of the centripetal force and since we have here, in my view, no centripetal force, there should be no centrifugal force here."}\\
In the comments, the student describes the role of the centrifugal force as a counterforce to the centripetal force. This implies, that the student does not apply the concept of the inertial centrifugal force and, instead, refers to the concept of the reactive centrifugal force which occurs as a reaction to a centripetal force.\\
And this is another example for the comments to a wrong answer in which the student assumes that only the Coriolis force is necessary to describe the trajectory of the airplane.\\
Instructor:\textit{``Please comment on your answer."}\\
M2: \textit{``The plane is deflected to the left from the point of view of $K^{\prime}$ and as it flies and the air friction is neglected, it has no contact with the ground and therefore no centrifugal force has to act which somehow has to keep it on a circular path and therefore I think that you can neglect that. But if now a person would rest in the center of $K^{\prime}$, he would see that the aircraft is apparently being deflected to the left. The plane actually flies straight ahead, but the disc on which the observer stands turns to the right. And therefore, seen in the rotating system, only the Coriolis force acts which would deflect the aircraft."}\\
Here, the student argues that the missing contact of the airplane to the rotating frame of reference is responsible for the description via the Coriolis force.\\
Instructor:\textit{``Please justify your answer of question Q1."}\\
H: \textit{``To describe the trajectory in $K^{\prime}$, both the Coriolis and the centrifugal force are needed. This is the case because, first, we have a velocity of the airplane in the rotating frame of reference. That's why we need a Coriolis force. And since there is a distance $r^{\prime}$ to the center of rotation, which is the origin of the coordinate system K', there must also be a centrifugal force."}\\ 
In the arguments, the student directly refers to the non-zero velocity $v^{\prime}$ of the object in $K^{\prime}$ in the equation of the Coriolis force $F_{Cor}$ (Eq. (\ref{eq:F_Cor})) and to the non-zero distance to the center of rotation $r^{\prime}$ which is a necessary component in the equation of of the centrifugal force $F_{Cen}$ (Eq. (\ref{eq:F_Cen})).\\
In question Q2, the distractor (e) has the highest probability among the incorrect answers. This corresponds to the answer that the water is sloshed opposite to the direction of the sum vector of Coriolis and centrifugal force. This answer would be correct if there were real forces acting on the glass. This is the reason for the answer of the student M1 who selected this answer:\\
Instructor:\textit{``Please tell us why you have selected this answer."}\\
M1: \textit{``The water spills out for sure, because of the inertia of the water, so it's just a question of how it spills out and I've decided to tick the answer (e), because of the idea that the water goes straight ahead and thus the direction of motion is precisely directed opposite to these forces. Because it does not matter to the water, whether it is in the rotating system or not."}\\
In this reasoning to question Q2, the student seems convinced of the fact that Coriolis and centrifugal force cause an effect in the stationary reference system $K$.\\ 
And this is the comment to the answer of the high achieving student H to question Q2.\\
Instructor:\textit{``Please give a reason to your answer to question Q2."}\\
H: \textit{``For the description of the trajectory, the Coriolis and centrifugal force are introduced and in K the glass makes a straightforward uniform motion. But since both are only apparent forces, they are only of relevance for the trajectory description in $K^{\prime}$ and do not really affect the glass in the reference system $K$, the water does not spill over. So, in this straight uniform motion, no force acts on the glass."}\\ 
In this reasoning, the student refers to the fictitious character of the Coriolis and centrifugal force and draws the correct solution by relating the uniform motion of the glass to the absence of forces in the stationary reference system $K$.\\
The reasoning of student M2 is similar to the one of H, therefore it is not displayed here. 
\subsection{Confidence levels affect visual focus}
\begin{figure}[ht!]
	\includegraphics[width=\linewidth]{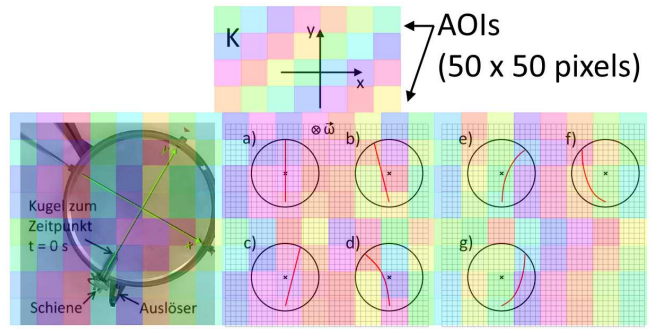}
	\caption{Locations of square-shaped AOIs ($50 \times 50$ pixels) during the $Predict$ and $Observe$ task.}
	\label{Fig4_AOIs}
\end{figure}

We were interested whether confidence ratings following the $Predict$ and the $Observe$ tasks have an influence on the visual attention of the students. For this purpose, we divided the students in two groups: the first group rated their confidence in this items with ``confident" or ``very confident", the students in the second group rated their confidence with ``unconfident" or ``very unconfident". Additionally, we designed a pattern of square-like Areas of Interest (AOIs) with a size of $50 \times 50$ pixels that covers all relevant areas (including the figure of the rotating disc, the coordinate system and the distractors) except the question text (see Fig. \ref{Fig4_AOIs}). 
Then, we compared the total visit duration and the size of the regions of attention. Table \ref{tab:Con_Pred} shows the visit duration and the number of AOIs $N_{AOIs}$ which received a high focus (i.e. a focus which is longer than the average focus of each student) of students who feel confident of their answer and those who are unconfident during the $predict$ task. Here, $N_{AOIs}$ is a measure of the size of the area of focus, i.e. it indicates the spatial spread of attention.\\
\begin{table}[ht]
	\centering
			\caption{Maximum, average, and total visit duration on AOIs in seconds during the $Predict$-task as well as the number of AOIs ($N_{AOIs}$) which exhibit a visit duration larger than the average one.}
		\label{tab:Con_Pred}
		\begin{tabular}{lcccccccc}
		\hline
		  &$~~~$& Confident &$~~~$& Unconfident &$~~~$& $d$ &$~~~$& $p$ \\ \hline 
		Max [s] &$~~~$& 3.38 &$~~~$& 7.33 &$~~~$& 1.21 &$~~~$& 0.005 \\ 
		Average [s] &$~~~$& 0.47 &$~~~$& 0.76 &$~~~$& 1.11 &$~~~$& 0.005 \\ 
		$N_{AOIs}$ &$~~~$& 14.29 &$~~~$& 13.86 &$~~~$& -0.07 &$~~~$& 0.82 \\ 
		Total [s]&$~~~$& 25.75 &$~~~$& 44.56 &$~~~$& 0.82 &$~~~$& 0.03 \\ \hline
		\end{tabular}
\end{table}
The analysis demonstrates that there is significant very large effect in the maximum visit duration and a significant large effect in the average and total visit duration between confident and unconfident students during the predict questions. \\
\begin{table}[ht]
	\centering
		\caption{Maximum, average, and total visit duration on AOIs during the $Observe$-task in seconds as well as the number of AOIs ($N_{AOIs}$) which exhibit a visit duration longer than the average one.}
		\label{tab:Con_Obs}
		\begin{tabular}{lcccccccc}
		\hline
		  &$~~~$& Confident &$~~~$& Unconfident &$~~~$& $d$ &$~~~$& $p$ \\ \hline 
		Max [s]&$~~~$& 2.74 &$~~~$& 5.68 &$~~~$& 0.68 &$~~~$& 0.16 \\
		Average [s]&$~~~$& 0.45 &$~~~$& 0.65 &$~~~$& 0.53 &$~~~$& 0.20 \\ 
		$N_{AOIs}$ &$~~~$& 8.03 &$~~~$& 11.77 &$~~~$& 0.79 &$~~~$& 0.06 \\ 
		Total [s]&$~~~$& 14.00 &$~~~$& 31.60 &$~~~$& 0.73 &$~~~$& 0.11 \\ \hline
		\end{tabular}
\end{table}
Table \ref{tab:Con_Obs} shows the visit duration and the number of AOIs which received a high attention. It is noticeable that the difference in the maximum, average, and total visit duration of confident and unconfident students is clearly reduced in comparison to the $Predict$-questions so that the effects are not significant during the $Observe$-questions. Furthermore, the results indicate that there is no significant difference in $N_{AOIs}$ between confident and unconfident students, which means that the studied area from where information is processed is similar between these two student groups.
\section{Discussion}
In this work we demonstrated how students understand concepts of rotating frames of reference and how they apply their knowledge to understand a standard lecture experiment in which they are supposed to report the trajectory of a sphere rolling over a rotating disc in a rotating and in a stationary coordinate system. \\
The presented study reveals a number of misconceptions in the field of rotating frames of reference which leads to a number of promising suggestions for future instructions of the topic. \\
The ``Confusion"-error, which is the error to think that there is a linear trajectory of the sphere in the rotating frame of reference or a curved trajectory of the sphere in the stationary coordinate system, was the most common error of students in the $Predict$ as well as in the $Observe$ items. This error could be successfully resolved via the instruction using a fundamental theoretical review of rotating frames of references, augmented photographs and six augmented videos. However, the results highlight the high difficulty of this topic for first-year physics students. This yields, for instance, the surprising observation that the ``Inversion"-error is neither corrected during the observation of the experiment nor during the instruction. This indicates that some errors require special attention which potentially could be realized via the implementation of cues \cite{cues} or via highlighting and discussing common errors of students in advance. For this reason, it is likely that a briefer instruction could fail to transfer the link between mathematical equations of the Coriolis and centrifugal force and their application to the trajectory of the sphere in a rotating and stationary frame of reference. \\
Furthermore, the item difficulty of the POE tasks prove the conceptual and perceptual complexity of the topic of rotating frames of references. Only one out of five physics students was able to report the observation of the trajectory of a sphere rolling over the disc correctly in a single choice question. This is significantly less than previous reports of POE interventions \cite{Crouch,Miller}. This implies that there was no obvious improvement which can unambiguously be attributed to the observation of the experimental demonstration, which was previously pointed out by Crouch et al. \cite{Crouch}. However, the detailed analysis of distractors of the single choice questions during POE in combination with the identification of different error types reveals the hidden benefits of lecture demonstration for learning. The probabilities of all error types show decreasing trends between $Predict$ and $Observe$ with the exception of the $inversion$ error. We could identify that, particularly false accounts for initial conditions can be corrected. This type of analysis clearly reveals the benefits of lecture demonstrations on learning about the outcome of the experiment. 
\subsection{Misconceptions related to Coriolis and centrifugal force}
The interviews as well as the distractor analysis of the POE items reveal prevailing misconceptions among first-semester physics students in the field of rotating frames of reference. Nearly half of the participants (42.8 \%) believe that the centrifugal force is only necessary to describe the trajectory of an object in a rotating coordinate system when there is a coupling of the object to the rotating system. This misconception is likely to be attributed to common instructional connections of the inertial centrifugal force and the reactive centrifugal force which occurs as a reaction to a Centripetal force, as for instance in a carousel. In the light of these results, we suggest to verbally discriminate between these two types of centrifugal forces. Apart from that, about one out of four physics students (23.8 \%) do not include the fictitious character of inertial forces in their arguments and rather argue that they have the same effect on objects as real forces. 
\subsection{Eye-tracking reveals confidence}
The Eye-Tracking analysis reveals a direct link between confidence ratings and visit duration on AOIs during the items of the $Predict$ phase. Students which are confident of their answer spent significantly less time on the AOIs than unconfident students. Despite this fact, unconfident students distribute their attention on a similar-sized area. This observation is an interesting extension to previous results and interpretations of long visit durations. For instance, Palinko et al. report that high visual attention on relevant areas is related to a high mental effort \cite{Palinko}. As a consequence, the visit duration has also been used as a measure for (intrinsic or extrinsic) cognitive load within the framework of the Cognitive Load Theory \cite{Mayer}. During the $Observe$ phase, there is no difference in the average or total visit duration between confident and unconfident students. The disappearance of the aforementioned relation between visit duration and confidence ratings in the $Observe$ part might be attributed to the fact that the students have seen the exact same questions already during the $Predict$ phase and have naturally less time-on-task since the content of the page is already partially familiar to the students. This interpretation is supported by an overall decrease of visit durations.\\ 
Furthermore, we observe that students with low confidence levels and high visit durations in the $Predict$ items distribute their focus on a similar-sized area as confident students. This seems to indicate that unconfident students tried longer to extract the same amount of information as confident students. In the theoretical framework of Rau \cite{Rau1}, the author points towards necessary prerequisites for learning using multiple visual representations. To identify and extract relevant information from a visual representation such as a graph, photograph, or schematic, students need visual representational understanding, which refers to the conceptual knowledge of how a visual representation depicts information. In order to relate the information from two different visual representations, as it is required in several parts of this study, the students need connectional understanding of two or more representations. This knowledge refers to the ability to identify relevant similarities between the representation and to know about conventions for interpreting and combining the information from multiple representations \cite{Rau1}. Embedding our results in this framework, it seems that unconfident students seem to try to develop visual and/or connectional understanding of the representations displayed in the $Predict$-items. 
\section{Conclusion}
In this study we tested the conceptual learning of physics students during a POE task on rotating frames of reference. The students had significant difficulties in predicting and observing the correct trajectory of a sphere (total score of approx. 20 \%) rolling over a rotating disc in a stationary and a rotating coordinate system $K$ and $K^{\prime}$. Primarily, the low score can be attributed to the misconception of a confusion of the effects of inertial forces in $K$ and $K^{\prime}$. Additionally, we found that some misconceptions even withstood the instruction. Students who initially predicted that the sphere is deflected to the opposite side on the disc (in respect to the actual trajectory) kept this conception during the $Observe$ and $Explain$ part (``Inversion" error). This emphasizes the need for additional instructional support in this topic for instance via cues which highlight essential information.\\
Furthermore, the results indicate that after the instruction nearly half of the students answered that a centrifugal force will only be necessary if there is a coupling between the object and the rotating system. In comparison, the misconception that an object shows a reaction to inertial forces in the same way as they do to real objects only persists in one quarter of the students.\\
Within the POE task, the eye tracking analysis in combination with confidence ratings showed that unconfident students spent significantly more time extracting information than confident students. This finding demonstrates the cognitive activation particularly of unconfident students during the $Predict$ phase. In contrast to previous reports we found that passive observations of experiments, in fact, stimulate conceptual learning in a detailed distractor analysis which is not reflected in an increase of total scores. At this point we cannot judge the importance of this non-obvious learning behavior and additional research is necessary. In this way the results assist to understand conceptual learning during POE tasks.
\section{Acknowledgment}
This work is funded by the Federal Ministry of Education and Research (BMBF; project: VorleXung; support code: 16DHL1001). The authors are responsible for the content of this contribution.
\section{Appendix}
\subsection{Theoretical background on rotating frames of reference}
When an observer examines motion of an object moving uniformly in a stationary frame of reference (SFR) from a rotating frame of reference (RFR), the trajectory appears to be curved in comparison to a trajectory which a stationary observer (SO) would report. For instance, if an object moved uniformly in a SFR, it would display a curved trajectory for a rotating observer (RO). The theoretical description of the trajectory in a RFR requires the introduction of the centrifugal and the Coriolis force. They are ``virtual forces" which means that Newton's third law of motion (\textit{action = reaction}) does not hold for them. In other words, both forces are not the result of an interaction between two bodies but the consequence of the motion within a RFR. If either one, the inertial centrifugal or the Coriolis force, acts on a body, there is no reaction from that body in the opposite direction. They are also called "inertial forces" which emphasizes the fact that the forces are caused by the inertia of the moving object. Typical examples include the motion of clouds observed from the earth or a thrown ball observed from a person sitting in a rotating merry-go-round \cite{Demtroeder}.\\
The velocity $\vec v\ ^{\prime}$ of an object in a RFR which rotates with a constant angular velocity $\vec \omega$ is given by the sum of the velocity $\vec v$ of the object with position $\vec r$ in the SFR and the negative tangential velocity $-\vec \omega \times \vec r$ in the RFR :
\begin{equation} \label{eq:vel}
\vec v\ ^{\prime}=\vec v - \vec \omega \times \vec r.
\end{equation} 
The derivative $\frac{\mathrm{d}\vec v\ ^{\prime}}{\mathrm{d}t}$ leads to the acceleration of the object in the RFR \cite{Demtroeder}:
\begin{equation} \label{eq:accel}
\vec a\ ^{\prime}=\vec a+\vec \omega \times \left(\vec r \times \vec\omega \right)+2 \left(\vec v\ ^{\prime} \times \vec\omega \right).
\end{equation}
This equation shows the necessity of introducing additional terms apart from the acceleration $\vec a$ in the SFR for the mathematical description of the determination of $\vec{a}\ ^{\prime}$. The second term in Eq. (\ref{eq:accel}) corresponds to the inertial centrifugal acceleration and points radially outwards from the axis of rotation. The third term is called the Coriolis acceleration which is perpendicular to the velocity vector $\vec v\ ^{\prime}$ in the plane of motion. \\
From Eq. (\ref{eq:accel}) the terms for the Coriolis force follow:
\begin{equation} \label{eq:F_Cor}
\vec F_{Cor}=2 m\left(\vec v\ ^{\prime} \times \vec\omega \right),
\end{equation}
and the equation for the centrifugal force:
\begin{equation} \label{eq:F_Cen}
\vec F_{Cen}= m\vec \omega \times \left(\vec r \times \vec\omega \right).
\end{equation}
In both equations $m$ denotes the mass of the object.
\subsection{Questions for student interviews}
The following questions were used during the student interviews:
\begin{enumerate}
\item[Q1: ]  Now imagine that, instead of the sphere, an airplane flies straight and uniformly from the center horizontally outward over a rotating disk. The disc rotates at constant angular velocity $\vec \omega$. The coordinate system $K^{\prime}$ has its origin in the center of the disk and also rotates at the velocity $\vec \omega$. Please neglect air friction. Which forces are necessary to describe the trajectory of the airplane in $K^{\prime}$? Please justify your answer. \\Answers: a) With the help of Coriolis and centrifugal force, b) Only with the help of the centrifugal force, c) Only with the help of the Coriolis force, d) No forces need to be introduced.
\begin{figure}[ht!]
	\includegraphics[width=0.6\linewidth]{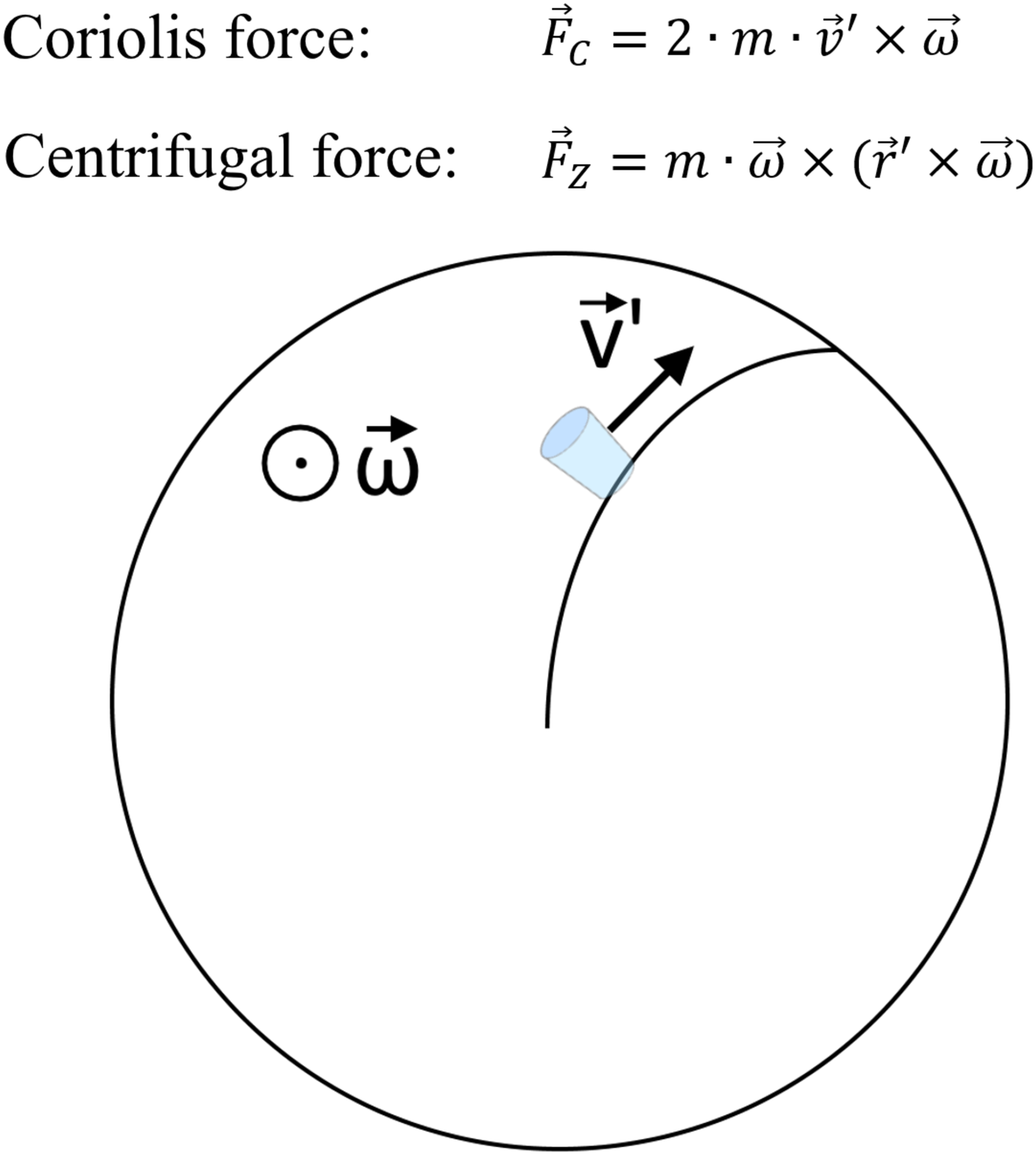}
	\caption{Figure of the moving water glass which corresponds to interview question Q2.}
	\label{Fig6_water}
\end{figure}
\item[Q2: ] A glass is completely filled with water and moves in a straight line and uniformly in a stationary coordinate system $K$ without friction over a rotating disk. The coordinate system $K^{\prime}$ rotates just like the disc with the constant angular velocity $\vec \omega$. The figure below shows the trajectory of the glass in $K^{\prime}$ (see Fig. \ref{Fig6_water}). To describe the motion in $K^{\prime}$, a Coriolis force and a centrifugal force are introduced. Is the water sloshing over the edge? If so, in which direction? Please justify your answer. \\Answers: a) Yes, in the direction of the centrifugal force, b) Yes, opposite to the direction of motion, c) Yes, in the direction of the Coriolis force, d) Yes, opposite to the direction of the Coriolis force, e) Yes, opposite to the sum vector of the Coriolis and centrifugal force, f) No.
\end{enumerate}

\end{document}